\documentclass{PoS}
\usepackage{epsfig,dcolumn}
\usepackage{graphicx}
\usepackage{color}
\usepackage{amsmath}
\usepackage{amsfonts}
\usepackage{amssymb}
\usepackage{graphicx}
\usepackage{type1cm}
\usepackage{eso-pic}

\newcommand{\be}{\begin{equation}}
\newcommand{\ee}{\end{equation}}

\usepackage{bm} 

\title{Calculation of resonances from $K \pi$ scattering}

\ShortTitle{Calculation of resonances from $K \pi$ scattering}

\author{\speaker{A.~ Rodas}\\
        Departamento de F\'isica Te\'orica II and UPARCOS, Universidad Complutense de Madrid, 28040 Madrid, Spain\\
        E-mail: \email{arodas@ucm.es}}
\author{J.R.~Pelaez\\
        Departamento de F\'isica Te\'orica II and UPARCOS, Universidad Complutense de Madrid, 28040 Madrid, Spain}
\author{J.~ Ruiz de Elvira\\
Albert Einstein Center for Fundamental Physics, Institute for Theoretical Physics,
University of Bern, Sidlerstrasse 5, 3012 Bern, Switzerland}

\abstract{We present a determination of the mass, width and coupling of the strange resonances appearing in pion-kaon scattering below 1.8 GeV, namely the much debated $K^*_0(800)$ or $\kappa$, the scalar $K^*_0(1430)$, the $K^*(892)$ and $K^*(1410)$ vectors, the spin-two $K^*_2(1430)$ as well as the spin-three $K^*_3(1780)$. The parameters of each resonance are determined using a direct analytic continuation of the pion-kaon partial waves by means of Pad\'e approximants, thus avoiding any particular model description of their pole positions and residues, while taking into account the analytic requirements imposed by dispersion relations.}

\FullConference{XVII International Conference on Hadron Spectroscopy and Structure - Hadron2017\\
		25-29 September, 2017\\
		University of Salamanca, Salamanca, Spain}

\begin{document}

\section{Introduction}

A reliable determination of strange resonances 
is by itself relevant for hadron spectroscopy and their own classification in multiplets, 
as well as for our understanding of intermediate energy QCD and the low-energy regime through Chiral 
Perturbation Theory. In addition $\pi K$ scattering and the resonances that appear in it are also of interest
because most hadronic processes with net strangeness end up with at least a $\pi K$ pair
that contributes decisively to shape the whole amplitude through final 
state interactions. Moreover, the inelastic resonances are not reachable through single-channel dispersion relations, so that it is not possible to obtain a direct model independent calculation of their parameters.

Very often the analyses of these resonances have been made  in terms of crude models,
which make use of specific parameterizations like isobars, 
Breit-Wigner forms or modifications, which assume the existence of some simple background and a model dependent relation between the pole position and residue.
As a result, resonance  parameters suffer from large model dependencies and may be process dependent. 
Thus, the statistical uncertainties  in the resonance parameters should be accompanied by systematic errors that are usually ignored. 

For the above reasons there is a growing interest in methods based on analyticity properties 
to extract resonance pole parameters from data in a given energy domain.
They are based on several approaches: conformal expansions to exploit the maximum analyticity domain of the amplitude~\cite{Cherry:2000ut},
Laurent~\cite{Guo:2015daa}, Laurent-Pietarinen~\cite{Svarc:2013laa} expansions,
Pad\'e approximants~\cite{Masjuan:2013jha,Pelaez:2016klv}, or the rigorous dispersive approaches~\cite{Ananthanarayan:2000ht}, note that the latter are in practice limited to $\sqrt{s}\sim$1 GeV.
They all determine the pole position without assuming a particular model for the relation between the mass, width and residue.
In this sense they are model independent analytic continuations to the complex plane. 

These analytic methods require as input some data description. 
It has been recently shown~\cite{Pelaez:2016tgi} that in the case of $\pi K$ scattering data~\cite{Estabrooks:1977xe}, which are the source for several determinations of strange resonances, 
they do not satisfy well Forward Dispersion Relations (FDR) up to 1.8 GeV.
This means that in the process of extracting data by using models, they have become in conflict with causality.
Nevertheless, in~\cite{Pelaez:2016tgi} the data were refitted  constrained to satisfy those FDR and following~\cite{Perez:2015pea} a careful systematic and statistical error analysis was provided.
In~\cite{Pelaez:2016klv} we made use of the Pad\'e approximants method in order to extract the parameters of all resonances appearing in those waves.

\section{Method and results}

Using the parameterizations obtained in~\cite{Pelaez:2016tgi},  we have a set of equations that are compatible with FDRs in the real axis, so that we can use Pad\'e  approximants to perform the analytic continuation to the complex plane. 
The $P^N_M(s,s_0)=Q_N(s,s_0)/R_M(s,s_0)$  Pad\'e approximants  of a function $F(s)$ around the point $s_0$ is a rational function that satisfies $P^N_M(s,s_0)=F(s)+O((s-s_0)^{M+N+1})$, 
with  $Q_N(s,s_0)$ and $R_M(s,s_0)$ polynomials in $s$ of order $N$ and $M$, respectively. 	
In the case of one pole in the complex plane the formula reads
\vspace{-0.2cm}
\begin{equation}
P^N_1(s,s_0)=\sum^{N-1}_{k=0}{a_k(s-s_0)^k+\frac{a_N(s-s_0)^N}{1-\frac{a_{N+1}}{a_N}(s-s_0)}},
\end{equation}
\vspace{-0.2cm}
where $a_n=\frac{1}{n!}F^{(n)}(s_0)$, and the position and residue of the pole are
\begin{equation}
s_p^{N}=s_0+\frac{a_N}{a_{N+1}},~Z^{N}=-\frac{(a_N)^{N+2}}{(a_{N+1})^{N+1}}.
\label{eq:poleresidue}
\end{equation}
If there is more than one pole inside the disk of convergence of the approximant, one has to change the order of the polynomial $R_M(s,s_0)$, both the formula of the pole and the residue change, but the parameters of the approximant are still directly related to the Taylor expansion.

With this simple analytic continuation we can go to the next continuous Riemann sheet and find not only the elastic but also {\it inelastic} resonances. We define the position of the pole as $\sqrt{s_p}=M-i\Gamma/2$, where the sequence is truncated when the different values of the poles $s_p^{N}$ and $s_p^{N+1}$ are much smaller than the statistical uncertainty obtained by running a montecarlo  simulation for the fit parameters.

The systematic errors of each pole are calculated using different parameterizations fulfilling FDRs so that we take into account the different values of the derivatives produced by slight changes of the parameterizations in the real axis. For example, for the $K^*_0(800)$ or $\kappa$ resonance we have used three different parameterizations: Schenk-like or Chew-Mandelstam parameterization and a conformal mapping. All them reproduce the required analytic structures, having slightly different values on the real axis that produce deviations of the pole position, this reflects that a wide resonance like the $\kappa$ cannot be described through simple models due to the instability of the analytic continuation.

In the case of the $\kappa$ resonance, which is the lightest strange resonance, the calculation is compatible with the most rigorous dispersive result, showing the good agreement between both analytic methods. Our result is $\sqrt{s_p}=(670\pm 18)-i(295\pm 28)$ MeV, while the result estimated by the PDG~\cite{PDG} is $\sqrt{s_p}=(682\pm 29)-i(274\pm 12)$ MeV. The values obtained for the rest of the strange resonances appearing below 1.8 GeV are listed in Table~\ref{tab:resonances}.
\begin{figure}
\centering
\includegraphics[width=0.7\linewidth]{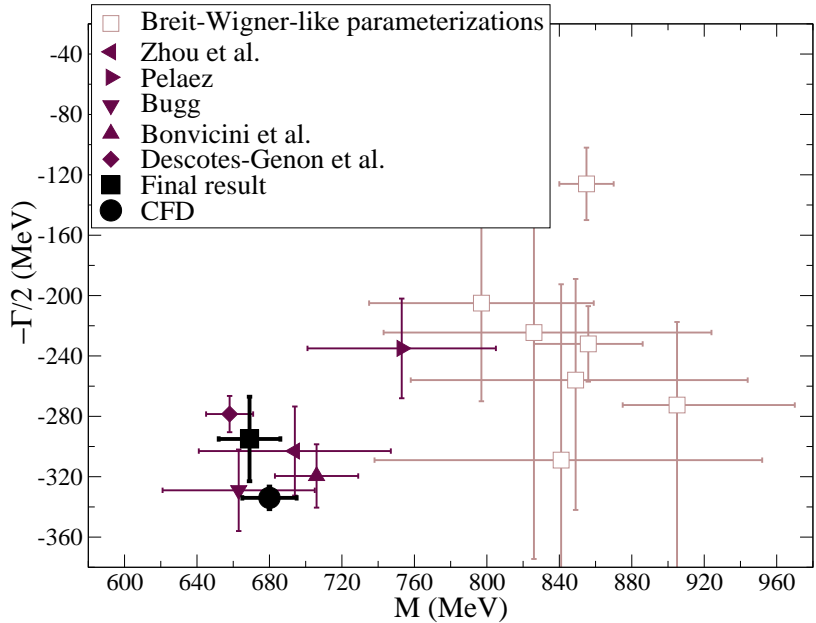}
\caption{\rm \label{fig:kappa} 
Final result for the $\kappa$ pole. Other references are taken from the RPP compilation~\cite{PDG}.}
\end{figure}

Note that even though our description of the $K^*(892)$ is compatible with the data in that region, the pole position we obtain (Table~\ref{tab:resonances}) is slightly different from the values listed in the PDG. The reason is that most of the values listed there are not the pole position parameters but the Breit-Wigner ones.

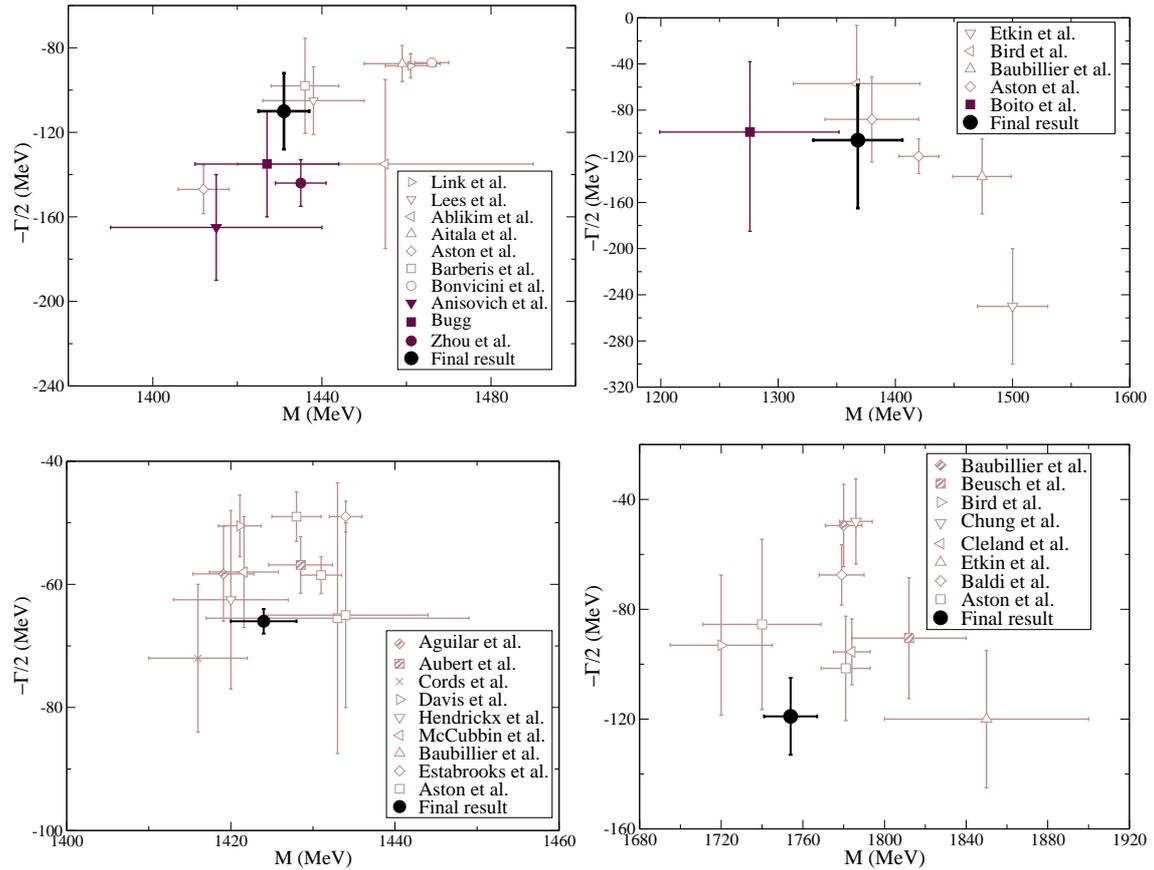
\begin{figure}
\centering
\centerline{\includegraphics[width=0.49\linewidth]{k0starfinal.eps} \includegraphics[width=0.49\linewidth]{k1starfinal.eps}}
\vspace{0.2cm}
\centerline{\includegraphics[width=0.49\linewidth]{k2starfinal.eps} \includegraphics[width=0.49\linewidth]{k3starfinal.eps}}
\label{fig:resonances}
\end{figure}

As explained before, this method is also suitable for the calculation of non-elastic resonances, which poles are not reachable through a dispersive approach, this is the case of the $K^*_0(1430)$, $K^*(1410)$, $K^*_2(1430)$ and $K^*_3(1780)$ resonances. In Fig.~\ref{fig:resonances}, we show the spread of results listed in the PDG for every resonance, while in Table~\ref{tab:resonances} we compare our results with the average given by the PDG.

In the case of the $K^*(1410)$, which branching ratio to $K \pi$ is 7$\%$, the errors are produced mostly due to the uncertainties of solutions of~\cite{Pelaez:2016klv}. The best result for the vector partial wave is incompatible with the result of Aston et al., which is the one used to determine the PDG values.

For high angular momentum ($l=2,3$)~ partial waves there is a remarkable spread of results, produced probably by the different models and Breit-Wigner-like descriptions, barrier factors, etc... . Instead, our approach avoids such a model dependent assumption, and it does not produce deviations from the pole position due to barrier factors or other kinematics included in those model dependent fits.

\begin{table}[h] 
\caption{Resonance parameters.} 
\centering 
\begin{tabular}{c r r } 
\hline\hline  
Resonance & Pad\'e $\sqrt{s_p}$ (MeV) & PDG $\sqrt{s_p}$ (MeV)\\ 
\hline 
$K^*_0(1430)$  & (1431$\pm$6)- i(110$\pm$19) & (1425$\pm$50)- i(135$\pm$40)\\
$K^*(892)$   & (892$\pm$1)-i(29$\pm$1) & (892$\pm$1)- i(25$\pm$1)\\
$K^*(1410)$  & (1368$\pm$38)-i(106$^{+48}_{-59}$) & (1421$\pm$9)- i(118$\pm$18)\\
$K^*_2(1430)$  & (1424$\pm$4)-i(66$\pm$2)& (1432$\pm$1)- i(55$\pm$3)\\
$K^*_3(1780)$  & (1754$\pm$13)-i(119$\pm$14) & (1776$\pm$7)- i(80$\pm$11)\\
\hline 
\end{tabular} 
\label{tab:resonances} 
\end{table} 
\vspace{-0.7cm}

\section{Summary}

Using the data parameterizations constrained to satisfy the dispersion relations obtained in~\cite{Pelaez:2016tgi} (see talk \cite{Pelaez:2017ppx} at Hadron 2017) we have calculated in~\cite{Pelaez:2016klv} the parameters of the strange resonances appearing up to 1.8 GeV thanks to the method of the Pad\'e approximants. The values obtained for the parameters of the resonances are in agreement with other works in the PDG, although our approach is based on a data analysis consistent with analyticity and makes use of a method that does not include any model to extract the parameters, providing a realistic estimate of systematic uncertainties.

Apart from the inelastic resonances we have also determined the mass, width and coupling to $K\pi$ for 
the conflictive $K_0^*(800)$ or $\kappa$ resonance, still not confirmed according to the PDG, while taking care of the systematic uncertainties for its parameters.

\vspace{-0.2cm}
\section{Acknowledgments}
Work supported by the Spanish Projects FPA2014-53375-C2-2, FPA2016-75654-C2-2-P and the Swiss National Science Foundation. 
A.~Rodas acknowledges the Universidad Complutense for a doctoral fellowship.

\end{document}